# SIGNCRYPTION SCHEME BASED ON SCHNORR DIGITAL SIGNATURE

Laura Savu

Department of Information Security, Faculty of Mathematics and Computer Science,
University of Bucharest, Bucharest, Romania
laura.savu@microsoft.com

## ABSTRACT

*This article presents a new signcryption scheme which is based on the Schnorr digital signature algorithm. The new scheme represents my personal contribution to signcryption area. I have been implemented the algorithm in a program and here are provided the steps of the algorithm, the results and some examples. The paper also contains the presentation of the original Signcryption scheme, based on ElGamal digital signature and discusses the practical applications of Signcryption in real life.*

The article is structured in seven parts, as follows. Signcryption and its properties definitions are contained in the first part. Also here, in introduction, are presented the practical applications of Signcryption in real life. In the second part is exposed the original signcryption primitive introduced by Youliang Zheng, which combines public key encryption and a derivation of ElGamal digital signature algorithm. Part three contains the presentation of the new syngcryption scheme, Schnorr Signcryption, as a result of the combination of public key encryption and Schnorr digital signature algorithm. The step-by-step implementation of the Schnorr Signcryption scheme in a source code program is reflected in the fourth part. Strating with the fifth part begins the analyze of the security models on Schnorr Signcryption. The two-users security model is presented in the sixth part and multi-user security model is presented in the seventh part. In each of this models there is exposed another classification for security, the insider security and the outsider security.

## 1. INTRODUCTION

Signcryption is the primitive that has been proposed by Youliang Zheng in 1997 and combines public key encryption with digital signature in a single logical step, obtaining a less cost for both communication and computation [1].
Data confidentiality and data integrity are two of the most important functions of modern cryptography. Confidentiality can be achieved using encryption algorithms or ciphers, whereas integrity can be provided by the use of authentication techniques. Encryption algorithms fall into one of two broad groups: private key encryption and public key encryption. Likewise, authentication techniques can be categorized by private key authentication algorithms and public key digital signatures.
While both private key encryption and private key authentication admit very fast computation with minimal message expansion, public key encryption and digital signatures generally require heavy computation, such as exponentiations involving very large integers, together with message expansion proportional to security parameters (such as the size of a large composite integer or the size of a large finite field).

Signcryption has the intention that the primitive should satisfy "Cost(Signature & Encryption) << Cost(Signature) + Cost(Encryption)." This inequality can be interpreted in a number of ways:

• A signcryption scheme should be more computationally efficient than a native combination of public-key encryption and digital signatures.

• A signcryption scheme should produce a signcryption "ciphertext" which is shorter than a naive combination of a public-key encryption ciphertext and a digital signature.

• A signcryption scheme should provide greater security guarantees and/or greater functionality than a native combination of public-key encryption and digital signatures [1].

More recently, the significance of signcryption in real-world applications has gained recognition by experts in data security. Since 2007, a technical committee within the International Organization for Standardization (ISO/IEC JTC 1/SC 27) has been developing an international standard for signcryption techniques [7].

The shared secret key between the parties makes possible an unlimited number of applications. Among these applications, one can first think of the following three:
- secure and authenticated key establishment,
- secure multicasting, and
- authenticated key recovery.

A number of signcryption-based security protocols have been proposed for aforementioned networks and similar environments. These include:
- secure ATM networks,
- secure routing in mobile ad hoc networks,
- secure voice over IP (VoIP) solutions,
- encrypted email authentication by firewalls,
- secure message transmission by proxy, and
- secure message transmission by proxy, and
- mobile grid web services.

The mobile ad hoc networks get subjected to security threats like other wireless networks. But due to their peer to peer approach and absence of infrastructural resources the mobile ad hoc networks cannot use strong cryptographic mechanisms as used by their other wireless counterparts. This led to the development of trust based methods as security solutions wherein a trusted node is relaxed from security checks when the trust value reaches to a particular limit. The trust methods are prone to security risks but have found their acceptance due to efficiency over computationally expensive and time consuming cryptographic methods. The major problem with the trust methods is the period during which trust is growing and is yet to reach the requisite threshold [10].

There are also various applications of signcryption in electronic commerce, where its security properties are very useful. Analyzing this security scheme from an application-oriented point of view, can be observed that a great amount of electronic commerce can take advantage of signcryption to provide efficient security solutions in the following areas:
- • electronic payment,
- • electronic toll collection system,
- • authenticated and secured transactions with smart cards, etc.

A related new public key primitive called Attribute-based encryption (ABE) attracted much attention recently. ABE has significant advantage over the traditional PKC primitives as it achieves flexible one-to-many encryption instead of one-to-one. ABE is envisioned an important tool for addressing the problem of secure and fine-grained data sharing and access control [9].

My personal contribution to the article is represented by the Schnorr Signcryption scheme which has been introduced here. Schnorr Signcryption scheme is made up of a combination between a public key encryption scheme and a digital signature scheme. On the base of the scheme that I present here stands the Schnorr digital signature. A Schnorr signature is a digital signature produced by the Schnorr signature algorithm. Its security is based on the intractability of certain discrete logarithm problems. It is considered the simplest digital signature scheme to be provably secure in a random oracle model. It is efficient and generates short signatures.

A signcryption scheme typically consists of five algorithms, Setup, KeyGenS, KeyGenR, Signcrypt, Unsigncrypt:
- Setup - takes as input a security parameter $1^k$ and outputs any common parameters *param* required by the signcryption schemes. This may include the security parameter $1^k$, the description of a group G and a generator g for that group, choices for hash functions or symmetric encryption schemes, etc.
- Key Generation S(Gen) - generates a pair of keys for the sender
- Key Generation R(Gen) - generates a pair of keys for the receiver
- Signcryption (SC) - is a probabilistic algorithm
- Unsigncryption (USC) - is a deterministic algorithm.

A signcryption scheme is a combination between a public key encryption algorithm and a digital signature scheme.

A public key encryption scheme consists of three polynomial-time algorithms (EncKeyGen, Encrypt, Decrypt).

**EncKeyGen** - Key generation is a probabilistic algorithm that takes as input a security parameter $1^k$ and outputs a key pair (skenc, pkenc), written (skenc, pkenc)R←EncKeyGen($1^k$). The public encryption key pkenc is widely distributed, while the private decryption key skenc should be kept secret. The public key defines a message m ∈ M and a ciphertext ∈ C.

**Encrypt** - Encryption is a probabilistic algorithm that takes a message m ∈ M and the public key pkenc as input and outputs a ciphertext C ∈ C, written C ←Encrypt(pkenc,m)

**Decrypt** - Decryption is a deterministic algorithm that takes a ciphertext C ∈ C and the private key skenc as input and outputs either a message m ∈M or the failure symbol ⊥, written m ← Decrypt(skenc,C).

## 2. ELGAMAL SIGNCRYPTION

ElGamal signcryption is the original signcryption scheme that has been introduced by Youliang Zheng in 1997. It is created on a derivation of ElGamal digital signature standard, combined with a public key encryption scheme.

Based on discrete algorithm problem, Signcryption cost is:
  58% less in average computation time
  70% less in message expansion

Here is the detailed presentation of the fifth algorithms that make up the ElGamal signcryption scheme.

1) Setup
  Signcryption parameters:
  p = a large prime number, public to all
  q = a large prime factor of p-1, public to all
  g = an integer with order q modulo p, in [1,… , p-1], public to all
  hash = a one-way hash function
  KH = a keyed one-way hash function = $KH_k(m)$ = hash(k, m)
  (E, D) = the algorithms which are used for encryption and decryption of a private key cipher.
  Alice sends a message to Bob.

2) KeyGen sender
  Alice has the pair of keys $(X_a, Y_a)$:

Xa = Alice's private key, chosen randomly from [1, .., q-1]

Ya = Alice's public key = $g^{Xa}$ mod p

3) KeyGen receiver

Bob has the pair of keys(Xb, Yb):

Xb = Bob's private key, chosen randomly from [1, .., q-1]

Yb = Bob's public key = $g^{Xb}$ mod p.

4) Signcryption

In order to signcrypt a message m to Bob, Alice has to accomplish the following operations:

Calculate

$k = hash(Yb^X) \bmod p$

Split k in k1 and k2 of appropriate length.

Calculate r = KHk2(m) = hash(h2, m)

Calculate s = x/(r+Xa) mod q, if SDSS1 is used

Calculate s = x/(1+Xa · r) mod q, if SDSS2 is used

Calculate c = Ek1(m) = the encryption of the message m with the key k1.

Alice sends to Bob the values (r, s, c).

5) Unsigncryption

In order to unsigncrypt a message from Alice, Bob has to accomplish the following operations:

Calculate k using r, s, g, p, Ya and Xb

$hash((Ya \cdot g^r)^{s \cdot Xb} \bmod p)$, if is used SDSS1

$hash((g \cdot Ya^r)^{s \cdot Xb} \bmod p)$, if is used SDSS2

Split k in k1 and k2 of appropriate length.

Calculate m using the decryption algorithm m = Dk1(c).

Accept m as a valid message only if KHk2(m) = r.

Using the two schemes SDSS1 and SDSS2, two signcryption schemes have been created, SCS1 and SCS2, respectively. The two signcryption schemes share the same communication overhead, (|hash(*)| + |q|). SCS1 involves one less modular multiplication in signcryption then SCS2, both have a similar computational cost for unsigncryption [1].

## 3. SCHNORR SIGNCRYPTION

A Schnorr signature is a digital signature produced by the Schnorr signature algorithm. Its security is based on the intractability of certain discrete logarithm problems. It is considered the simplest digital signature scheme to be provably secure in a random oracle model [3].

**Choosing parameters**

All users of the signature scheme agree on a group G with generator g of prime order q in which the discrete log problem is hard.

**Key generation**

- Choose a private signing key *x*.
- The public verification key is $y = g^x$.

*Signing*

To sign a message M:

- Choose a random k.
- Let $r = g^k$
- Let e = H(M || r), where || denotes concatenation and r is represented as a bit string. H is a cryptographic hash function $H : \{0,1\}^* \rightarrow \mathbb{Z}_q$.
- Let s = (k − xe).

The signature is the pair (s,e).

*Verifying*

- Let $r_v = g^s y^e$

- Let $e_v = H(M || r_v)$

If ev = e then the signature is verified.

*Demonstration of correctness*

It can be observed that ev = e if the signed message equals the verified message:
$r_v = g^s y^e = g^{k-xe} g^{xe} = g^k = r$, and hence $e_v = H(M || r_v) = H(M || r) = e$.

It has been considered that k < q and the assumption that the hash function is collision-resistant.

Public elements: G, g, q, y, s, e, r.

Private elements: k, x. [4]

A Schnorr Signcryption scheme is based on Schnorr digital signature algorithm. Here is the detailed presentation of the fifth algorithms that make up the Schnorr signcryption scheme.

1) Setup

Schnorr Signcryption parameters:

p = a large prime number, public to all
q = a large prime factor of p-1, public to all
g = an integer with order q modulo p, in [1,… , p-1], public to all
hash = a one-way hash function
KH = a keyed one-way hash function = KHk(m) = hash(k, m)
(E, D) = the algorithms which are used for encryption and decryption of a private key cipher.
Alice sends a message to Bob.

2) KeyGen sender

Alice has the pair of keys (Xa, Ya):
Xa = Alice's private key, chosen randomly from [1, .., q-1]
Ya = Alice's public key = $g^{-Xa}$ mod p

3) KeyGen receiver

Bob has the pair of keys(Xb, Yb):
Xb = Bob's private key, chosen randomly from [1, .., q-1]
Yb = Bob's public key = $g^{-Xb}$ mod p.

4) Signcryption

In order to signcrypt a message m to Bob, Alice has to accomplish the following operations:
Calculate
$k = hash(Yb^X) \; mod \; p$
Split k in k1 and k2 of appropriate length.
Calculate r = KHk2(m) = hash(h2, m)
Calculate s = x + (r * Xa) mod q
Calculate c = Ek1(m) = the encryption of the message m with the key k1.
Alice sends to Bob the values (r, s, c).

5) Unsigncryption

In order to unsigncrypt a message from Alice, Bob has to accomplish the following operations:
Calculate k using r, s, g, p, Ya and Xb
$k = hash((g^s * Ya^r)^{-Xb} \; mod \; p)$
Split k in k1 and k2 of appropriate length.
Calculate m using the decryption algorithm m = Dk1(c).
Accept m as a valid message only if KHk2(m) = r.

Analyzing the two presented signcryption schemes, it can be observed that in case of Shnorr signcryption the computation of s, which is s = x + (r * Xa) mod q, is less consuming comparing with the formula used in ElGamal algorithm, where s is s = x/(r+Xa) mod q.

Another difference is on the level of unsigncryption step as k is computing differently, using this formula for Schnorr $k = hash((g^s * Ya^r)^{-Xb} \; mod \; p)$ and this formula for ElGamal $k = hash((Ya \cdot g^r)^{s \cdot Xb} \; mod \; p)$.

# 4. IMPLEMENTATION OF SCHNORR SIGNCRYPTION SCHEME

I created a source code program that verifies my algorithm. Executing this program I could generate examples. The step-by-step implementation of the algorithm is as follows:

1)Calculate Ya and Yb
```
double powA = Math.Pow(g, xA);
int pow_intA = Convert.ToInt32(powA);
int invA = modInverse(pow_intA, p);
```

2)Calculate k
```
int yB = Convert.ToInt32(textBox11.Text);
int x = Convert.ToInt32(textBox18.Text);
int p = Convert.ToInt32(textBox4.Text);
string cheie = (BigInteger.ModPow(yB, x, p)).ToString();
```

3)Calculate hash(k)
```
 string HashDeCheie = _calculateHash(cheie);
 textBox13.Text = HashDeCheie;
```

4)Split k in two keys k1 and k2 with the same lenght
```
byte[] k = Convert.FromBase64String(textBox13.Text);
byte[] k1 = new byte[k.Length/2];
byte[] k2 = new byte[k.Length - k.Length / 2];
Buffer.BlockCopy(k, 0, k1, 0, k.Length/2);
Buffer.BlockCopy(k, k.Length / 2, k2, 0, k.Length - k.Length / 2);
byte[] test = new byte[k.Length];
k1.CopyTo(test, 0);
k2.CopyTo(test, k1.Length);
```

5)Calculate r using k2;  r = hash (k2, m)
```
BigInteger p = BigInteger.Parse(textBox4.Text);
System.Text.ASCIIEncoding encoding = new System.Text.ASCIIEncoding();
byte[] keyByte = encoding.GetBytes(key);
HMACSHA1 hmacsha1 = new HMACSHA1(keyByte);
byte[] messageBytes =encoding.GetBytes(message);
byte[] hashmessage = hmacsha1.ComputeHash(messageBytes);
```

6)Calculate r using k2; transform the value obtained from hash in base 10
```
textBox19.Text = fn16to10(textBox15.Text).ToIntString();
```

7)Calculate the modulo p of the number obtained in base 10
```
BigInteger nr=BigInteger.Parse(textBox19.Text);
BigInteger p = BigInteger.Parse(textBox4.Text);
BigInteger  rest = 0;
BigInteger.DivRem(nr, p, out rest);
```

8)Calculate s
```
BigInteger q = Convert.ToInt32(textBox5.Text);
BigInteger  r = Convert.ToInt32(textBox20.Text);
BigInteger XA = Convert.ToInt32(textBox9.Text);
BigInteger X = Convert.ToInt32(textBox18.Text);
BigInteger prod = BigInteger.Multiply(r, XA);
BigInteger sum = X + prod;
BigInteger rest;
BigInteger.DivRem(sum, q, out rest);
```

9) Encrypt m using the k1

10) Calculate k
```
BigInteger rez2 = BigInteger.Pow(rez1, XB);
BigInteger invK = modInverseBI(rez2, p);
```

Here is provided an example from the execution of the program on small numbers.
Example:
p = 23, q = 11, g = 2, X=3
XA=4 => YA=13
XB=5 => YB=18
k = 13 => hash(k) = vTB6PsMp4Qos/4+4dICCPaEU+PQ=
k1 = vTB6PsMp4Qos/w==
k2 = j7h0gII9oRT49A==
hash(k2, m) = E2726583242AB5CCE58AE1151DB126208F17932F
hash(k2,m) in base 10 = 1292783042124763369608714420962730428414981280559
(hash(k2,m) in base 10) mod p = 3
s mod q = x+(r*Xa)  mod q = 4
Unsigncrypt k  = 13

## 5. SECURITY MODELS FOR SCHNORR SIGNCRYPTION

The first attempt to produce security models for signcryption was given by Steinfeld and Zheng [6].
A family of security models for signcryption in both two-user and multi-user settings was presented by An [5] in their work on signcryption schemes built from black-box signature and encryption schemes.

Defining the security of signcryption in the public-key setting is more involved than the corresponding task in the symmetric setting [8] due to the asymmetric nature of the former. The asymmetry of keys makes a difference in the notions of both authenticity and privacy on two major fronts which are addressed in this chapter.

The first difference for Schnorr signcryption is that the security of the signcryption needs to be defined in the multi-user setting, where issues with users' identities need to be addressed. On the other hand, authenticated encryption in the symmetric setting can be fully defined in a much simpler two-user setting.

The case of Schnorr settings not only makes a difference in the multiuser and two-user settings but also makes a difference in the adversary's position depending on its knowledge of the keys. There are two definitions for security of signcryption depending on whether the adversary is an "outsider" (a third party who only knows the public information) or "insider" (a legal user of the network, either the sender or the receiver, or someone that knows the secret key of either the sender or the receiver). In the first case the security model is named "outsider security" and in the latter "insider security".

## 6. TWO-USERS SECURITY MODEL

In the symmetric setting, there is only one specific pair of users who
(1) share a single key;
(2) trust each other;
(3) "know who they are";
(4) only care about being protected from "the rest of the world."

In contrast, in Schnorr signcryption setting, each user independently publishes its public keys, after which it can send/receive messages to/from any other user. In particular, (1) each user should have an explicit identity (associated with its public key); (2) each signcryption has to explicitly contain the (presumed) identities of the sender S and the receiver R; (3) each user should be protected from every other user.

The security goal is to provide both authenticity and privacy of communicated data. In the symmetric setting, since the sender and the receiver share the same secret key, the only security model that makes sense is one in which the adversary is modeled as a third party or an outsider who does not know the shared secret key. For Schnorr signcryption setting, the sender and the receiver do not share the same secret key but each has his/her own secret key. Due to this asymmetry of the secret keys, the data needs to be protected not only from an outsider but also from an insider who is a legal user of the system (the sender or the receiver themselves or someone who knows either the sender's secret key or the receiver's secret key) [2].

### 1) Outsider Security Model

The adversary A has the public information represented by the sender's public key and the receiver's public key (pkS, pkR). He also has oracle access to the functionalities of both the sender and the receiver. Specifically, it can mount a chosen message attack on the sender by asking the sender to produce a Schnorr signcryption C of an arbitrary message m. In other words, A has access to the signcryption oracle. Similarly, it can mount a chosen ciphertext attack on the receiver by giving the receiver any candidate Schnorr signcryption C and receiving back the message m (where m could be ⊥), which means that A has access to the unsigncryption oracle. A cannot by itself run either the signcryption or the unsigncryption oracles due to the lack of corresponding secret keys skS and skR.

### 2) Insider Security Model

In case of the insider security model, the attacker is given one of the private keys of the users. If the attacker is the receiver, he has the private key of the receiver and the signcryption scheme prevents a receiver from forging a signcryption ciphertext that purports to be from the sender. This is a necessary condition if non-repudiation is to be achieved. In the other situation, if the attacker is the sender, he has the private key of the sender and the signcryption scheme prevents a sender from deciphering a signcryption ciphertext that has previously been produced. This means that the Schnorr signcryption scheme protects the confidentiality of messages even if the sender's private key is subsequently leaked to an attacker.

## 7. MULTI-USER SECURITY MODEL

A central difference between the multi-user model and the two-user models is the extra power of the adversary. In the multi-user model, the attacker may choose receiver (resp. sender) public keys when accessing the attacked users' signcryption (resp. unsigncryption) oracles. For signcryption schemes that share some functionality between the signature and the encryption components, such as are the case for Zheng's Signcryption scheme and Schnorr Signcryption scheme, the extra power of the adversary in the multi-user model may be much more significant, and a careful case-by-case analysis is required to establish security of such schemes in the multi-user model.

As in the two-user setting, the multi-user setting also has two types of models depending on the identity of the attacker: an insider model and an outsider model.

### 1) Outsider Security Model

The outsider model assumes an attack by an entity who does not know either the sender or receiver's secret keys. In the multi-user model, this confidentiality notion of signcryption is specially termed "indistinguishability of signcryptext against chosen ciphertext attack with access to 'flexible' signcryption/unsigncryption oracles (FSO/FUOIND- CCA2)." The indistinguishability of signcryptext (abbreviated by "IND") here means that there is no polynomial-time adversary that can learn any information about the plaintext from the signcryptext except for its length. In the chosen ciphertext attack for signcryption, it is assumed that an adversary has access to two oracles that perform signcryption and unsigncryption. The oracles are flexible in the sense that the adversary can freely choose the public keys with which those oracles perform Schnorr signcryption and Schnorr unsigncryption.

Let A = (A1,A2) be a two-stage adversary trying to break the confidentiality of messages between the (fixed) sender S and the (fixed) receiver R.

1. The Setup algorithm is run and the resulting common parameters, denoted by param, is sent to any interested parties including S, R, and A1.

2. The KeyGenS and KeyGenR algorithms are run to generate S and R's public/private key pairs, denoted by (skS, pkS) and (skR, pkR), respectively. The public keys (pkS, pkR) are given to A1.

3. A1 submits a series of signcryption and unsigncryption queries. Each signcryption query consists of a pair (pk,m) where pk is a receiver's public key generated by A1 and m is a message. On receiving this, the signcryption oracle computes a signcryptext C ← Signcrypt(param, skS, pk,m) and returns it to A1. Each unsigncryption query consists of a pair (pk',C) where pk' denotes a sender's public key generated by A1 at will and C is a signcryptext. On receiving this, the unsigncryption oracle performs unsigncryption by computing
Unsigncrypt(param, pk', skR,C) and returns the result to A1.

4. A1 outputs a pair of equal-length plaintexts (m0,m1) and a state string α. On receiving this, the Schnorr signcryption oracle picks b ←R {0, 1} at random, computes a target signcryptext C□ ←Signcrypt(param, skS, pkR,mb), and runs A2 on input (C□, α).

5. A2 submits a number of signcryption/unsigncryption queries as A1 did in Step 3. A restriction here is that A2 is not allowed to query (pkS,C□) to the unsigncryption oracle.

6. A2 outputs its guess b' ∈ {0, 1} for the value of b chosen in Step 4. A is said to win the game if b' = b.

### 2) Insider Security Model

Unlike the outsider setting where the attacker only knows the public keys of the attacked pair of users S and R, the insider model deals with the setting where an attacker, knowing the secret key of the sender S, tries to decrypt the Schnorr signcryptexts sent by that sender. In order to give the attacker as much power as possible, it is provided the sender's key pair.

1. The Setup algorithm is run and the resulting common parameters, denoted by param, is sent to any interested parties including S, R, and A1.

2. The key generation algorithm is run just once, to generate the attacked receiver's key pair (skR, pkR), and pkR is given to A1.

3. A has access to R's unsigncryption oracle, but not to the signcryption oracle.

4. A1 outputs an attacked sender's key pair (skS, pkS), in addition to (m0,m1, α). The key skS is used to produce the challenge signcryptext C∗ as in the outsider model.

5. A has access to R's unsigncryption oracle, but not to the signcryption oracle.

## ACKNOWLEDGEMENTS

This paper presents a new Signcryption scheme which is based on Schnorr digital signature algorithm. This scheme is named Schnorr Signcryption and it implements in a single logical step both public key encryption and digital signature, offering less costs as using these two cryptographic functions individually.

**Authors**

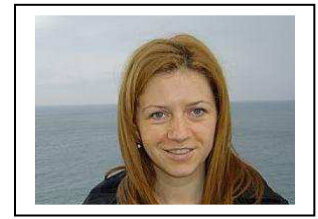

Short Biography

1996-2000 University of Bucharest, Faculty of Mathematics and Computer Science

2000-2011 Working in software development

Currently working for Microsoft and in the 3rd year as a Phd Candidate in Information Security